# Spin interference in silicon one-dimensional rings

Nikolay T. Bagraev,[a*] Nikolay G. Galkin,[a] Wolfgang Gehlhoff,[b] Leonid E. Klyachkin,[a] Anna M. Malyarenko,[a] Ivan A. Shelykh[a]

[a]*A.F. Ioffe Physico-Technical Institute, 194021, St.Petersburg, Russia*

[b]*Institut für Festkörperphysik, Technische Universität Berlin, D-10623 Berlin, Germany*

**Abstract**

We present the first findings of the spin transistor effect caused by the Rashba gate-controlled ring embedded in the p-type self-assembled silicon quantum well that is prepared on the Si (100) surface using the planar diffusion CVD technology. The coherence and phase sensitivity of the spin-dependent transport of holes are studied by varying the value of the external magnetic field and the gate voltage that are perpendicular to the plane of the double-slit ring. Firstly, the quantum scatterers connected to two one-dimensional leads and the quantum point contact inserted in the one of the arms of the double-slit ring are shown to define the amplitude and the phase of the Aharonov-Bohm and the Aharonov-Casher conductance oscillations. Secondly, the amplitude and phase sensitivity of the *"0.7∗$2e^2/h$)"* feature of the hole quantum conductance staircase revealed by the quantum point contact inserted are found to result from the interplay of the spontaneous spin polarization and the Rashba spin-orbit interaction.

*Keywords:* Spin interference, Self-assembled silicon quantum wells, Spin-orbit interaction, Aharonov-Bohm double-slit ring

## Introduction

The spin-correlated transport in low-dimensional systems was in focus of both theoretical and experimental activity in the last decade [1, 2]. The studies of the Rashba spin-orbit interaction (SOI) that results from the structure inversion asymmetry in mesoscopic nanostructures have specifically attracted much of the efforts [3]. The variations in the Rashba SOI value appeared to be a basis of spintronic devices such as the spin-interference device that is able to demonstrate the characteristics of the spin field-effect transistor (FET) even without ferromagnetic electrodes and external magnetic field [4]. This device shown schematically in Fig. 1 represents the Aharonov-Bohm (AB) ring covered by the gate electrode, which in addition to the geometrical Berry phase provides the phase shift between the transmission amplitudes for the particles moving in the clockwise and anticlockwise direction.

## Methods

Here we study the effects of the Rashba SOI on the amplitude and the transmission phase shift (TPS) of the *"0.7∗$2e^2/h$)"* feature in the quantum conductance staircase that is revealed by the quantum point contact (QPC), which is inserted within one of the arms of the one-dimensional ring inside the p-type self-assembled silicon quantum well (SQW) prepared on the n-type Si (100) surface (Fig. 1). The parameters of the SQW that contains the high mobility 2D gas of holes were defined by the SIMS, STM, cyclotron resonance (CR) and EPR methods. The one-dimensional ring embedded in SQW, R=2500 nm, contains the source and drain constrictions that represent quantum point contacts. The quantum conductance staircase caused by the QPC inserted within one of the arms of the one-dimensional ring was controlled by varying the split-gate voltage.

---

[*] Corresponding author. Tel.: +007-812-2479315; fax: +007-812-2471017; e-mail: impurity.dipole@mail.ioffe.ru

## Results

The amplitude of the *"0.7∗2e²/h)"* feature fixed by the split-gate voltage is shown to exhibit the Aharonov-Casher (AC) conductance oscillations that are due to the Rashba SOI by varying the gate voltage applied to the p$^+$n junction (Figs. 2a and 2c). The phase variations of the AB and AC oscillations observed seem to be caused by the scattering on the QPCs inside the double-slit ring which is analyzed in frameworks of the scattering matrix formalism [5]. The changes in the amplitude of the *"0.7∗2e²/h)"* feature up to the *0.7∗e²/h)* value under low gate voltage appear to result from the spontaneous spin polarization of heavy holes in the SQW that is created by decreasing of their concentration (Fig. 2b). The Fano resonance structure is also evidence of the spin polarization of holes inside the QPC inserted in the double-slit ring. Besides, the phase shift of the AB oscillations measured at the fixed amplitude of the *"0.7∗2e²/h)"* feature of the quantum conductance staircase is found to be changed by the saturation of the electrically-detected NMR of the $^{29}$Si nuclei thereby verifying directly the spin polarization of holes inside the QPC. This phase shift appeared to result from the Overhauser shift that is due to the regular magnetic field created by the $^{29}$Si nuclei.

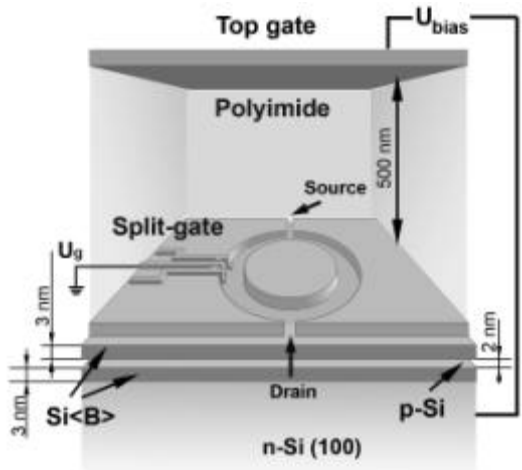

Fig. 1. Schematic diagram of the device that demonstrates a perspective view of the p-type silicon quantum well located between the δ - barriers heavily doped with boron on the n-type Si (100) surface, the top gate that is able to control the sheet density of holes and the depletion regions created by split-gate method, which indicate the double-slit ring with QPC inserted in one of its arms.

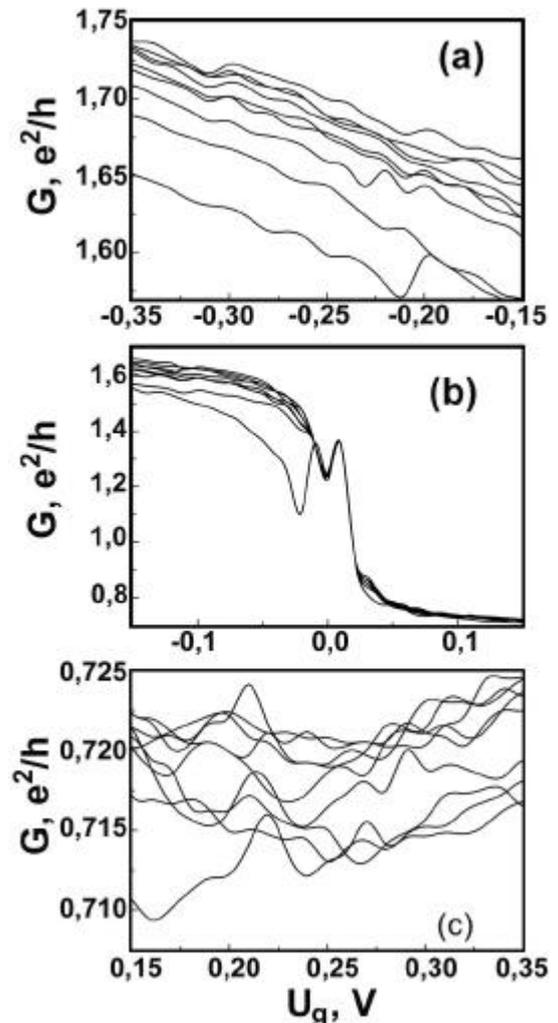

Fig. 2. The AC conductance oscillations (a, c) that attend the changes in the amplitude of the *"0.7∗2e²/h)"* feature (b) in the double-slit ring with the QPC inserted in one of its arms. The external magnetic field value was changed from 0.05 mT (bottom) to 0.5 mT (top) in 0.05 mT step.